**Intelligent Configuration of Integrated Microwave Photonic Filter Featuring Self-Stabilization and Programmable Response**

*Yutong Shi, Yuan Yu*, Yifan Liu, Kaixiang Cao, Mengmeng Deng, Fangzheng Zhang, Hailong Zhou, and Xinliang Zhang**

Y. T. Shi, Prof. Y. Yu, Dr. Y. F. Liu, K. X. Cao, M. M. Deng, Prof. H. L. Zhou, Prof. X. L. Zhang
Wuhan National Laboratory for Optoelectronics, Huazhong University of Science and Technology, Wuhan 430074, China
School of Optical and Electronic Information, Huazhong University of Science and Technology, Wuhan, 430074, China
E-mail: yuan_yu@hust.edu.cn; xlzhang@mail.hust.edu.cn

Prof. Y. Yu, Prof. X. L. Zhang
Optics Valley Laboratory, Wuhan, 430074, China
Dr. Y. F. Liu
National Information Optoelectronics Innovation Center, China Information and Communication Technologies Group Corporation, Youkeyuan Road 88, Wuhan 430074, China
Prof. F. Z. Zhang
National Key Laboratory of Microwave Photonics, Nanjing University of Aeronautics and Astronautics, Nanjing 210016, China

Integrated microwave photonic filters (IMPFs) have emerged as promising candidates for advanced microwave systems owing to their distinctive combination of wide operational bandwidth, flexibility, and compact size. Nevertheless, the complex and time-consuming manual manipulation of IMPFs remains a significant impediment to their widespread applications. Here, to the best of our knowledge, we experimentally demonstrate the first intelligent configuration of IMPF featuring wideband center frequency tunability, flexible bandwidth reconfigurability, self-stabilization, and excellent channel equalization simultaneously. The configuration is enabled by our proposed universal hybrid collaboration strategy, which can fully unleash the hardware potential of the optical device, thus enabling comprehensive synergy of multiple properties. Results show that the center frequency of IMPF is tuned from 2 to 48 GHz, covering microwave S- to Ka-bands, and the bandwidth is reconfigured from 0.66 to 4.15 GHz, with a rejection ratio of up to 37.67 dB. The roll-off rate and shape factor reach as high as 17.50 dB/GHz and 0.78, respectively. Meanwhile, the maximum center frequency drift of IMPF over 3 hours is reduced from 11.950 to 0.051 GHz even without a thermo-electric cooler, indicating that the center frequency stability is enhanced by 234 times. The passband shape of the IMPF can be dynamically adjusted to equalize frequency-dependent fading, achieving up to 2.42 dB of intra-channel fading compensation. Our work highlights the potential of IMPFs based on intelligent configuration, unlocking new avenues for practical applications of microwave photonic signal processing.

## 1. Introduction

Traditional radio frequency (RF) electronic technologies, constrained by limited sampling rates and component bandwidths, increasingly fail to meet the demands of next-generation communication systems, which require large bandwidth, flexible reconfigurability, excellent programmability, low cost, low consumption, and electromagnetic immunity[1,2,3,4]. In contrast, microwave photonics offers a breakthrough pathway by leveraging its inherent advantages, including ultra-wide bandwidth, high flexibility, low loss, and electromagnetic immunity[4,5,6]. As fundamental components, microwave photonic filters are used to select target signals from background noise or reject disturbing noise from signals. Advances in photonic integrated circuit (PIC) technology have driven the growing adoption of integrated microwave photonic filters (IMPFs)[7,8,9,10,11,12], which offer significant advantages in terms of size, weight, power consumption, and cost[4,8]. Currently, IMPFs have been demonstrated on various platforms, including silicon-on-insulator (SOI)[12,13,14,15,16,17,18,19], silicon nitride[20,21,22,23], indium phosphide[9,24,25], lithium niobate[26,27], and chalcogenide glass[28,29,30]. For example, a highly reconfigurable IMPF achieves monolithic integration of modulator, photodetector, and passive devices on the SOI platform[13]. Meanwhile, an ultra-wideband tunable IMPF leverages the TFLN's ultrahigh electro-optic bandwidth and low loss to enable impressive channel selection for giga bit per second data rates in high-frequency wireless communication[27].

Compared with electronic filters, IMPFs exhibit superior tunability and reconfigurability, which drive their broad application prospects in wireless communications[8,30,31], sensor networks[32], and radar systems[4,10]. IMPFs require tailored configurations for various scenarios. For example, optoelectronic oscillators (OEOs) require IMPFs with a narrow bandwidth[33,34], and radar systems demand IMPFs with a high rejection ratio (RR), wideband operation, and flexible tunability[19,35,36]. With the development of integrating multidimensional functions into a single system, reconfiguring IMPF responses according to diverse requirements significantly enhances economic efficiency.

Although the flexibility of IMPFs has been demonstrated[12,14,20,26], most reported reconfigurable IMPFs primarily depend on cumbersome manual manipulation for frequency tuning and bandwidth reconfiguration, which presents significant challenges for achieving real-time, dynamic, and self-adaptive configuration for complex applications. Many reconfigurable IMPFs utilizing various techniques have been demonstrated, such as Mach-Zehnder interferometer (MZI)-assisted microring resonators (MRRs)[13,37,38], cascaded MRRs[21,26,30,39], coupled resonator optical waveguides (CROWs)[15,40,41,42], and programmable two-dimensional (2D) mesh network topologies[12,20,43]. These IMPFs exhibit superior performance in key properties, including RR, tunable range of center frequency, reconfigurable range of full width at half-maximum (FWHM) bandwidth, shape factor (SF), and multifunctional integration. Nevertheless, these designs often encounter performance trade-offs due to insufficient configuration. Furthermore, their sustained and stable operation, as well as resistance to random environmental disturbances, have not been sufficiently addressed. Overall, traditional manual manipulation and offline design methods fail to meet the requirements for synergy of all properties, dynamic configuration, and sustained and stable operation of IMPFs, which significantly hinder their application prospects. Therefore, developing a supporting intelligent configuration framework is essential for advancing IMPF toward practical wireless applications.

In this work, we propose a universal multi-objective and multi-stage hybrid collaboration (MOMS-HC) strategy to implement intelligent configuration of IMPFs. The strategy combines global-local exploration, robust multi-objective equilibrium, efficient self-optimization, and environmental robustness empowerment, and fully unleashes the hardware potential of IMPFs. Utilizing the MOMS-HC strategy, we implement, to the best of our knowledge, the first intelligently configured bandpass IMPF, whose hardware is based on a 4th-order coupled resonator optical waveguide (CROW) on a commercial SOI wafer. Experimental results show that the IMPF attains unprecedented synergy of almost all properties, including a tuning range of 2-48 GHz covering microwave S- to Ka-bands, an FWHM bandwidth reconfiguration range of 0.66-4.15 GHz, a roll-off rate of up to 17.50 dB/GHz, an in-band ripple of below 0.91 dB, an RR of up to 37.67 dB, and an SF of up to 0.78. Notably, with the MOMS-HC strategy, the maximum center frequency drift of IMPF is only 0.051 GHz over 3 hours. Compared to the free-running case, the frequency stability is enhanced by 234 times. Results also indicate that the MOMS-HC strategy enables the IMPF to quickly recover to its target operating state after significant disturbances. Additionally, the IMPF-based equalization of frequency-dependent fading demonstrates up to 2.42 dB of compensation for intra-channel fading in a microwave photonic system by reshaping the IMPF passband. The proposed intelligent configuration method provides vital support for practical IMPF applications and is expected to accelerate the deployment of high-performance integrated optical devices.

## 2. Results

### 2.1 Principle of the intelligent configuration

The principle of the IMPF based on intelligent configuration is shown in Figure 1a. At first, the microwave signals are converted to optical signals after electrical-optical (E/O) conversion, which is realized by modulating the continuous wave (CW) light emitted by a laser source (LS). Then, the converted microwave photonic signal is processed by an integrated optical processing unit (IOPU). Here, the IOPU is used to implement the filtering function. Following the processing, the microwave signal is extracted through optical-electrical (O/E) conversion. The resulting microwave signal is partially output, while the rest is fed back to the intelligent configuration module, while the other portion is fed back to the intelligent configuration module. This module subsequently calculates the control signals based on the configuration targets. After calculation, the output control signals are applied to the IOPU, thereby enabling its intelligent reconfiguration. The integrated microwave photonic filtering is fundamentally implemented based on the optical-to-electrical mapping technique, and the performance of optical devices directly determines the quality of microwave signal processing. Therefore, the design and configuration of the IOPU are crucial to IMPF.

To meet the requirements of cognitive radio and complex RF tasks, we should consider two fundamental problems in realizing intelligent microwave photonic signal processing. The first problem is how to rapidly configure numerous adjustable parameters to ensure that the IMPF converges to the desired response, thus realizing software-defined hardware. The challenges of this problem include the vast parameter space resulting from complex structures and the inter-coupling between multiple adjustable parameters. Consequently, traditional manual manipulation is prone to falling into local optima, resulting in unsatisfactory frequency response properties. By contrast, the intelligent configuration can be enabled by an appropriate configuration strategy to achieve programmable responses, making the IMPF suitable for diversified scenarios, as shown in Figure 1b. Another critical issue is how to ensure the long-term stability of the IMPF, specifically how to suppress the performance degradation caused by environmental disturbances, such as temperature fluctuations[44], mechanical vibrations, and power supply noise[45,46]. In practical scenarios, the frequency response of the IMPF is highly susceptible to shape degradation and frequency drift over time, making it particularly important to construct a self-stabilizing mechanism for compensating response degradation, as shown in Figure 1c.

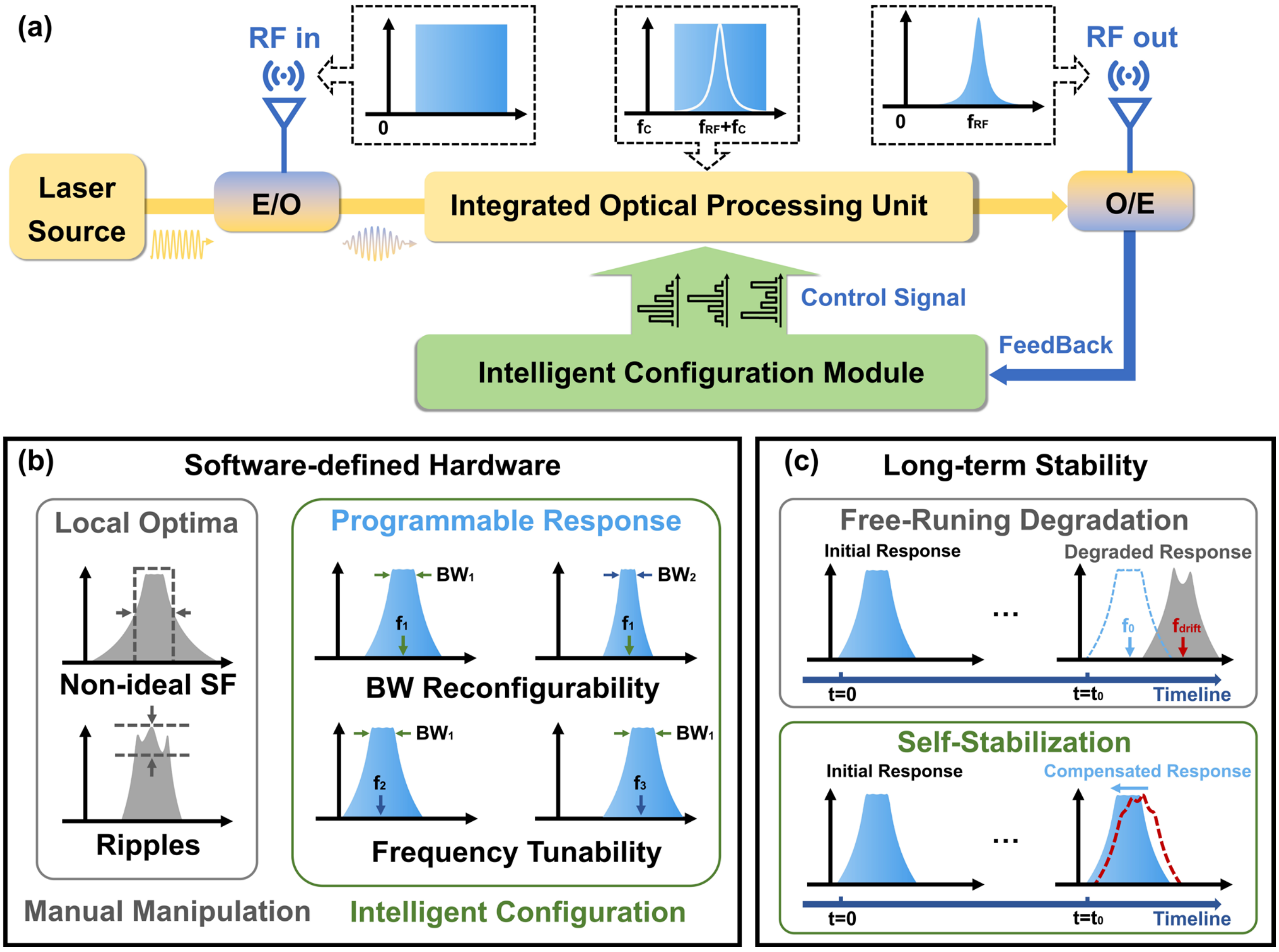

**Figure 1.** Intelligent configuration of the IMPF. (a) The principle of the integrated microwave photonic filter based on intelligent configuration. (b) Comparison between manual manipulation and intelligent configuration. (c) Comparison between time-dependent degradation and self-stabilization of IMPF.

To address the above two challenges, we propose transforming the configuration problem of IMPF into an optimization problem and solving it using advanced machine learning algorithms. The optimization objective and variables are the required frequency response and control voltage applied to the adjustable parameters. Meanwhile, to realize an IMPF with superior overall performance, we must prioritize the multi-objective equilibrium of the configuration strategy. Therefore, we propose the MOMS-HC strategy supporting the intelligent configuration of IMPF. The strategy, consisting of four stages, innovatively leverages the complementary strengths of particle swarm optimization (PSO)[47], genetic algorithm (GA)[48], and simulated annealing (SA)[49] through a staged collaboration mechanism. Specifically, PSO drives global parameter initialization, leveraging its swarm intelligence to accelerate convergence of IMPF toward the target response. GA executes precise configuration under multi-objective constraints by leveraging crossover and mutation operators. SA suppresses environmental disturbances through a dynamic fine-tuning search to ensure sustained and stable operation of IMPF. The task-driven specialization of each sub-algorithm at different stages governs the collaboration, enabling multi-algorithm hybrid execution and complementary strengths integration (see Methods for more details of the MOMS-HC strategy).

High-order CROWs exhibit excellent filtering characteristics such as high RR, high SF, low insertion loss, and flexible reconfiguration. However, due to their inherent structural complexity and strong inter-coupling between adjustable parameters in a compact size, dynamic configuration poses considerable challenges, making it difficult to unleash their hardware potential[15,40,42]. Therefore, we employ a 4th-order CROW as the IOPU and empower its configuration via the proposed MOMS-HC strategy to realize an intelligently configured IMPF with programmable response and self-stabilization.

### 2.2 Device Design and Experimental Setup

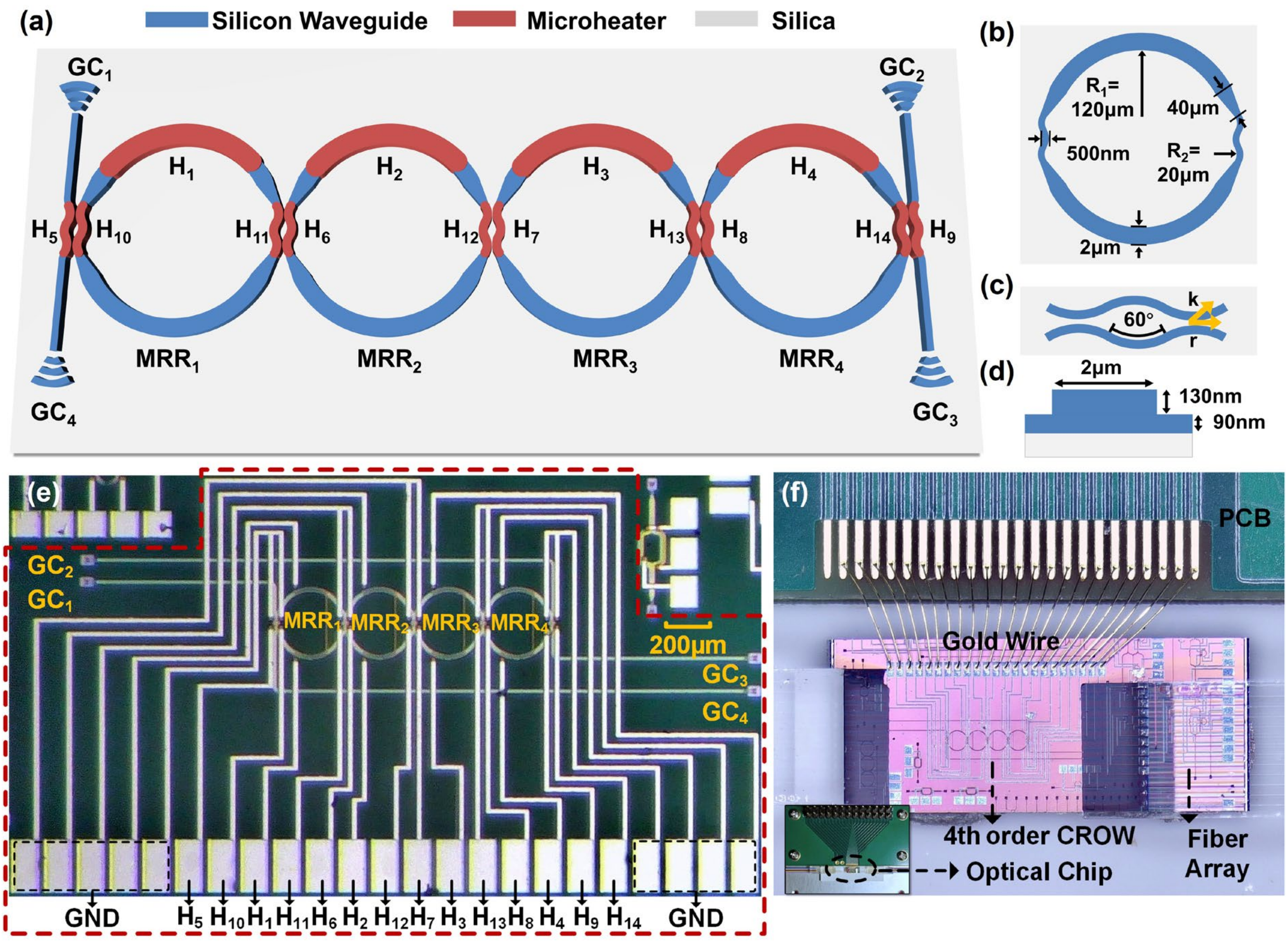

**Figure 2.** The details of the SOI-based device. (a) Schematic diagram of the designed 4th-order CROW to be implemented as the IMPF. (b) Schematic diagram of the MRR, in which the 500 nm-wide single-mode waveguide is connected to the 2-μm-wide multimode waveguide via a long linear adiabatic taper. (c) Schematic diagram of the MZI coupler, where r and k are the self-coupling and cross-coupling coefficients of DC, respectively. (d) Schematic diagram of the ridge waveguide with a ridge height of 130 nm. (e) The microscope image of the fabricated device. (f) The detailed photograph of the packaged chip. The inset shows the overview of the packaged module.

The 4th-order CROW is composed of four identical MRRs integrated with 14 titanium nitride (TiN) microheaters, as shown in Figure 2a. The detailed parameters of each MRR are shown in Figure 2b. MRR uses 2-μm-wide multimode waveguides to reduce sidewall scattering loss to achieve a high Q value for narrowband signal processing. The MRR coupling regions are implemented as MZI couplers consisting of two identical directional couplers (DCs) and two phase shifters. These couplers are designed as two single-mode waveguides with three 60° arcs and a bending radius of 20 μm, as illustrated in Figure 2c. The device employs ridge waveguides with a ridge height of 130 nm to reduce the waveguide loss, as shown in Figure 2d. The values of r and k should satisfy $r=k=1/\sqrt{2}$ in order to obtain the maximum tuning range of the equivalent coupling coefficients[39,41]. The FSR of the MRR is designed to be approximately 100 GHz (see Supplementary Note V for more details of the modulated signal).

Figure 2e shows the microscope image of the fabricated 4th-order CROW, which serves as the filtering device and is manufactured by Chongqing United Microelectronics Center (CUMEC). The device is fabricated based on an SOI wafer with a 220 nm-thick silicon layer and a 2 μm-thick buried oxide (BOX) layer. The four microheaters ($H_1$-$H_4$) deposited over individual MRR waveguides are used for tuning the resonant wavelength of each MRR. The remaining microheaters ($H_5$-$H_{14}$) deposited over the upper and lower arms of MZI couplers are used for adjusting equivalent coupling coefficients between coupled MRRs, as shown in Figures 2a and 2e. In practical operation, only one microheater ($H_5$-$H_9$) of the adjustable MZI coupler actively controls coupling coefficients, and the others ($H_{10}$-$H_{14}$) are used for backup. These microheaters enable refractive index modulation in waveguide structures by changing local temperature[20]. The measured linear resistances of the nine microheaters $H_1$-$H_9$ are 1.5491, 1.5532, 1.6022, 1.5848, 0.2109, 0.2001, 0.2184, 0.2361, and 0.2222 kΩ, respectively. To ensure the stability of the optical and electrical ports during experimental tests, the on-chip device is optically and electrically packaged, as shown in Figure 2f. All pads fabricated on the chip are connected to DuPont connectors on a printed circuit board (PCB) via gold wires. The four grating couplers (GCs) are coupled and packaged with a standard fiber array (FA). When the microwave photonic signal is coupled into the device through $GC_1$, the signal will be processed and then coupled out of the device through $GC_3$. The measured insertion loss of the device is 11.5 dB, including GC's single-end coupling loss of approximately 4 dB and CROW's inherent loss of approximately 3.5 dB.

Figure 3 illustrates the experimental setup for dynamically configuring the frequency response of the IMPF based on our proposed MOMS-HC strategy. A CW light at 1550 nm generated by a laser source (LS, Koheras BasiK E15) with an optical power of 15 dBm is divided into two branches by an optical coupler ($OC_1$) with a 7:3 splitting ratio. In the upper branch, 70% of the CW light is utilized for microwave signal modulation. The state of polarization (SOP) of the optical carrier is adjusted by a polarization controller ($PC_1$) to be aligned with the principal polarization axis of the phase modulator (PM, Covega Mach-40). The microwave signal emitted by a vector network analyzer (VNA, Anritsu MS4647B) is modulated onto the optical carrier via PM. An optical bandpass filter (OBPF, Alnair BVF-300CL) is used to eliminate the -1st-order sideband, generating a single-sideband (SSB) signal (see Supplementary Note IV for more details of the modulated signal). Then, the SSB signal is amplified by an erbium-doped fiber amplifier (EDFA, WZEDFA-SO-L-CW40) and processed by the device under test (DUT) after SOP adjustment via $PC_2$. The +1st-order sideband signal processed by the DUT is combined with the remaining 30% CW light in the lower branch via $OC_2$. The combined optical signal is then injected into a photodetector (PD, SHF 47100A) for O/E conversion to obtain the microwave signal. The microwave signal is amplified by a broadband electrical amplifier (EA, SHF S804B) before being fed into the VNA for measurement. $PC_3$ is used to adjust the SOP of the optical carrier in the lower branch to maximize the power of the beat frequency signal, and a variable optical attenuator (VOA) is used to adjust the optical power fed into the PD.

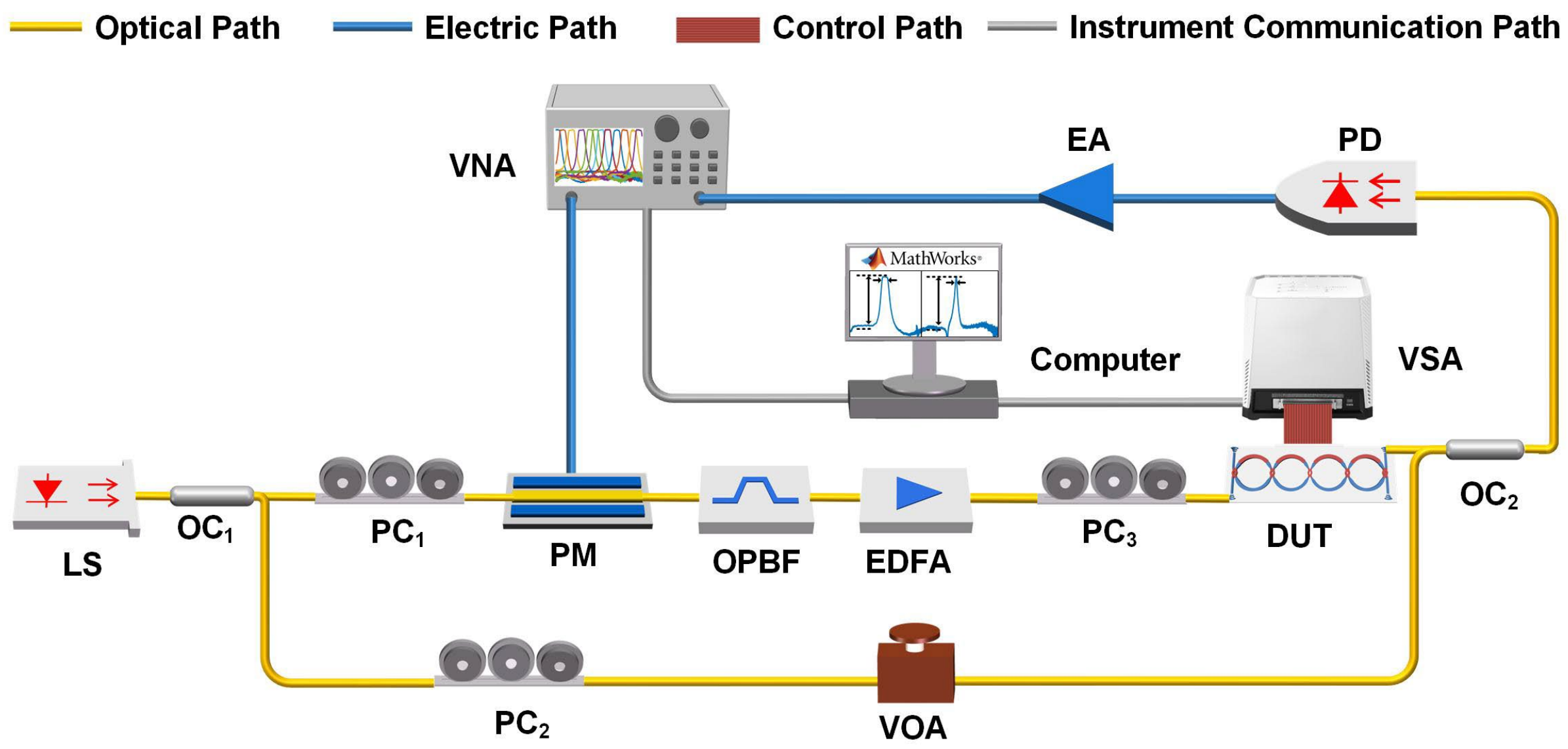

**Figure 3.** Schematic diagram of the experimental setup. Experimental setup for configuring the IMPF based on our proposed MOMS-HC strategy. LS laser source, OC optical coupler, PC polarization controller, VNA vector network analyzer, EA electrical amplifier, PM phase modulator, OBPF optical bandpass filter, EDFA erbium-doped fiber amplifier, DUT device under test, VSA voltage source array, PD photodetector.

Considering the extensive measurement range of the optical spectrum analyzer (OSA), we first conduct preliminary tests in the optical domain and execute the intelligent configuration of the DUT before configuring the IMPF. Based on the MOMS-HC strategy, an optical filter with an RR of 18.93 dB and an FWHM bandwidth of 0.018 nm (2.25 GHz) is realized (see Supplementary Note V for more details).

In brief, the intelligent configuration module consists of three components: controller, sensor, and actuator. The VNA functions as a sensor in the system, monitoring the operation state of the controlled device and relaying the data back to the computer, which serves as the central controller. The computer formulates the optimization problem via programming on the MATLAB platform and transmits the required control signals to the voltage source array (VSA, MCVS6400-A), which acts as the actuator. Both the sensor and the actuator are connected and controlled by the same computer.

### 2.3 Tuning and reconfiguring the response of IMPF

In traditional manual manipulation of the IMPF, operators usually rely on accumulated experience and subjective judgment, making it difficult to systematically explore the optimization parameter space and prone to getting stuck in local optima (see Supplementary Note III for more details). In contrast, the MOMS-HC strategy enables synergy of all properties and fully unleashes the performance potential of its hardware. We experimentally demonstrate the intelligent configuration process of the programmable IMPF with software-defined response (see Supplementary Note VI for more details).

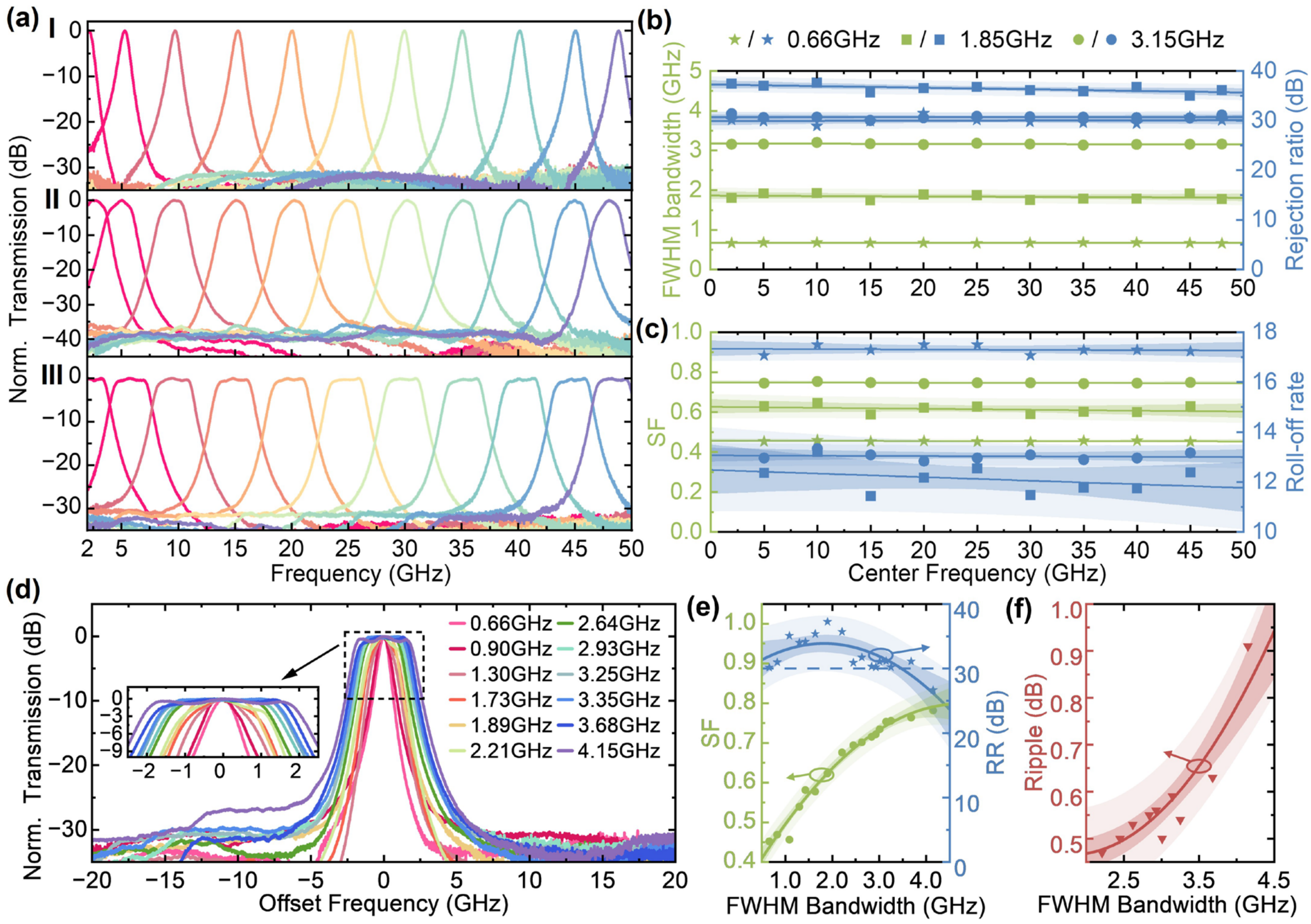

**Figure 4.** Tuning and reconfiguring the IMPF based on the proposed MOMS-HC strategy. (a) Tuning the center frequency of IMPF with different bandwidths, I 0.66GHz, II 1.85GHz, and III 3.15 GHz. (b)

Characterizing the variation of the FWHM bandwidth and RR during the frequency tuning. (c) Characterizing the variation of the SF and roll-off rate during the frequency tuning. (d) Reconfiguring the bandwidth of IMPF from 0.66 to 4.15 GHz. (e) Measured SF and RR with different FWHM bandwidths. (f) Measured in-band ripple with different FWHM bandwidths. The curves show polynomial fitting results. The dark shaded area indicates the 95% confidence interval, while the light shaded area represents the 95% prediction interval. The normalization is performed by taking the peak of the respective frequency response as the reference.

The powerful tunability and reconfigurability are essential capabilities of the IMPF. Therefore, we then evaluate its consistency while tuning the center frequency from 2 to 48 GHz with different target bandwidths to demonstrate the excellent performance of the intelligently configured IMPF.

Figure 4a illustrates the measured center frequency tuning results of the IMPF with bandwidths of 0.66, 1.85, and 3.15 GHz, respectively. From the measured results, we can draw the following conclusions. Firstly, for the narrow bandwidth operation (0.66 ± 0.02 GHz), the IMPF exhibits a maximum RR of 31.47 dB with a roll-off rate up to 17.50 dB/GHz, as shown in Figure 4a I. This high roll-off rate improves the rejection of out-of-band interference while preserving precise frequency selectivity, a crucial aspect of high-performance filtering. Secondly, for the medium bandwidth operation (1.85 ± 0.10 GHz), the maximum RR reaches 37.67 dB with a roll-off rate up to 13.21 dB/GHz, delivering balanced selectivity for multi-channel communication environments, as shown in Figure 4a II. Finally, for the wide bandwidth operation (3.15 ± 0.03 GHz), the IMPF maintains an in-band ripple of less than 0.61 dB, as shown in Figure 4a III. Meanwhile, the RR, roll-off rate, and SF are as high as 31.42 dB, 13.33 dB/GHz, and 0.75, respectively. The signal-to-noise ratio degradation caused by the insertion loss and inherent noise of devices in the microwave photonic system limits the RR of the IMPF.

Figures 4b and 4c illustrate the variation of FWHM bandwidth, RR, SF, and roll-off rate during the frequency tuning under different bandwidth operations (0.66, 1.85, and 3.15 GHz), respectively. We can see that the variations of FWHM bandwidth, SF, and RR during the center frequency tuning are less than ±0.10 GHz, ±0.03, and ±1.00 dB, respectively. The results demonstrate that the performance of the IMPF maintains excellent consistency across the whole operational band, representing a key advantage of the IMPF.

Meanwhile, the FWHM bandwidth of the IMPF can be continuously reconfigured from 0.66 to 4.15 GHz by jointly configuring the control voltages via the MOMS-HC strategy, and the SF varied between 0.45 and 0.78, as shown in Figures 4d and 4e, respectively. When the bandwidth is reconfigured from 0.66 to 3.68 GHz, the in-band ripple stays below 0.63 dB, and the RR stays above 30.14 dB, as shown in Figures 4e and 4f, respectively. When the bandwidth of the IMPF is reconfigured to 4.15 GHz, the RR drops to 26.66 dB, and the in-band ripple increases to 0.91 dB, as shown by the dark purple curve in Figure 4d. The degradation in overall performance primarily stems from the limited adjustable range of the equivalent coupling coefficient based on the MZI coupler, resulting in a trade-off between the RR and in-band ripple. The frequency response with a bandwidth of 4.15 GHz indicates that the 4th-order CROW is approaching its hardware-imposed performance limit. The result also demonstrates the capability of the configuration framework based on the MOMS-HC strategy to fully unleash the hardware potential of the optical device.

**2.4 Self-stabilization and self-recovery**

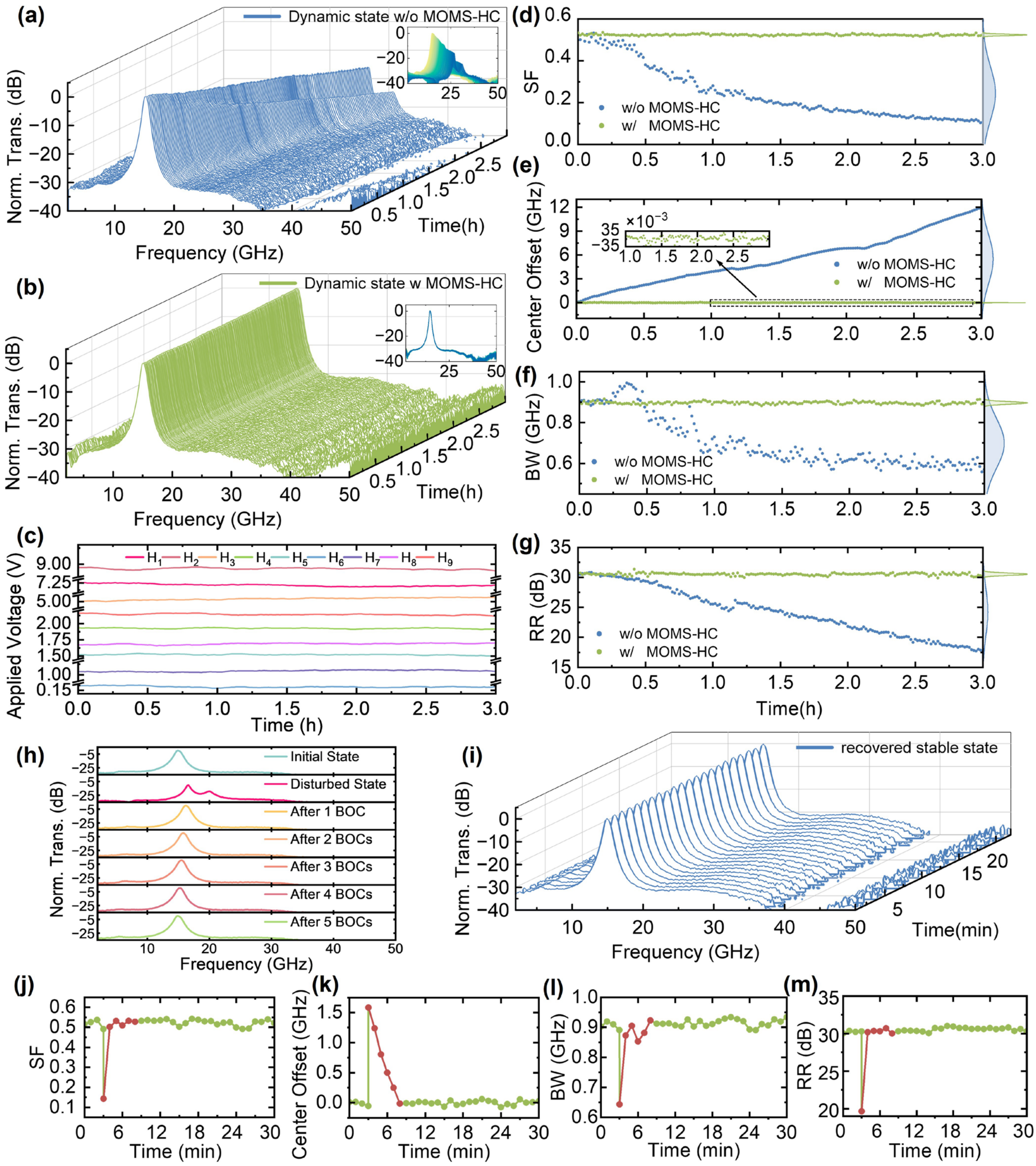

**Figure 5.** Self-stabilization and self-recovery of the intelligently configured IMPF. (a) The waterfall diagram showing the amplitude frequency response of the IMPF without the MOMS-HC configuration over 3 hours.

(b) The waterfall diagram showing the amplitude frequency response of the IMPF with MOMS-HC configuration over 3 hours. The inset shows the "eye-like" diagram generated by overlapping the captured frequency responses. The gradient color scale from light to dark green indicates the evolution of the frequency response over time. The frequency response of IMPF is normalized using the initial state as the reference. (c) Real-time dynamic adjustment of the nine voltages applied to the microheaters during IMPF operation with MOMS-HC on. (d)–(g) Time-dependent variations of (d) SF, (e) center frequency offset relative to the target value of 15.000 GHz, (f) bandwidth, and (g) RR during the 3-hour stability test, corresponding marginal plots show the normal statistical distribution of various properties. (h) The amplitude-frequency response of the IMPF in its initial operating state, after experiencing the disturbance, and during the subsequent self-recovery process. (i) The waterfall diagram showing the amplitude frequency responses of the IMPF in the subsequent 22 minutes after recovery, with the RR and roll-off rate above 30 dB and 15 dB/GHz, respectively. (j)-(m) Time-dependent variations of (k) SF, (l) center frequency offset, (m) bandwidth, and (n) RR. The red dots and curves in the figure indicate the recovery process.

The self-stabilization and sustained operation capability of the IMPF are critical for a practical system. Under manual manipulation, the absence of dynamic adjustment and stabilization mechanisms makes it challenging to maintain sustained and stable operation of the IMPF. In contrast, the MOMS-HC strategy enables dynamic self-configuration of adjustable parameters during IMPF operation to effectively suppress the negative effects of environmental disturbances. The elaborate design of the SA algorithm aims to provide real-time voltage compensation while minimizing temperature fluctuations as much as possible (see Supplementary Note I for more details of SA).

Before evaluating the self-stabilization of the IMPF, we first configure its various properties as follows. The center frequency and FWHM bandwidth are set to 15.000 GHz and 0.90 GHz, and the SF and RR are set to exceed 0.50 and 30.00 dB, respectively. The above state serves as the initial state for all subsequent evaluations in this section. Next, we let the IMPF operate in the laboratory environment for three hours with the MOMS-HC configuration on and off, respectively. Considering the response speed of sensors and actuators (e.g., VNA, VSA, and TiN microheaters), the basic operating cycle (BOC) is set to 60 seconds to balance operational stability, performance monitoring accuracy, and dynamic adjustment capability (see Supplementary Note VII for more details). When the MOMS-HC configuration is off, the IMPF operates normally within each BOC without executing additional configurations. When the MOMS-HC configuration is on, SA obtains the current amplitude frequency response of the IMPF in real-time during each BOC, monitoring degradation of properties such as center frequency, SF, RR, and bandwidth. Any degradation exceeding the threshold will trigger the intelligent configuration process, and the configuration is constrained to be completed within a BOC. At the end of each BOC, the current amplitude frequency response of the IMPF will be captured. The waterfall diagrams indicating the time-dependent variation of the IMPF's frequency responses, without and with the MOMS-HC configuration, are shown as Figures 5a and 5b, respectively. The inset shows the "eye-like" diagram generated by overlapping the captured frequency responses. The gradient color scale from light to dark green indicates the frequency response's evolution over time. We can observe that the waterfall diagram shows obvious irregularity when the MOMS-HC is off, indicating significant performance degradation of the IMPF over the 3 hours. The degradation is primarily caused by environmental disturbances, such as temperature fluctuations[44], mechanical vibrations, and accumulated power supply noise[45,46]. In contrast, when the MOMS-HC is on, the waterfall diagram distinctly exhibits sustained and stable dynamic characteristics of the IMPF. The nine voltages applied to the microheaters are dynamically adjusted in real time during IMPF operation, as shown in Figure 5c. The average electrical power of the nine control ports is 33.62, 51.35, 15.88, 2.35, 10.97, 0.21, 5.15, 12.08, and 20.67 mW, respectively. The total power consumption required to maintain IMPF's stability is approximately 152.26 mW.

Table 1. Comparison of IMPF with MOMS-HC on and off

| Property | | BW[b)] (GHz) | RR (dB) | Center (GHz) | SF |
|---|---|---|---|---|---|
| Target | | 0.90 | >30.00 | 15.000 | >0.50 |
| MOMS-HC off | Mean | 0.70 | 24.18 | 20.508 | 0.24 |
| | Min | 0.56 | 17.59 | 15.091 | 0.10 |
| | Max | 0.99 | 30.80 | 26.950 | 0.53 |
| | Rela.Std. Dev.[a)] | 17.553% | 16.367% | 15.048% | 51.708% |
| MOMS-HC on | Mean | 0.90 | 30.52 | 15.002 | 0.52 |
| | Min | 0.88 | 29.89 | 14.970 | 0.51 |
| | Max | 0.92 | 31.36 | 15.051 | 0.54 |
| | Rela.Std. Dev.[a)] | 0.682% | 0.648% | 0.085% | 0.667% |

[a)] Rela.Std. Dev.: relative standard deviation; [b)] BW: bandwidth.

Table 1 shows the comparison of IMPF with MOMS-HC on and off during the 3-hour stability test. When the MOMS-HC is on, the mean center frequency of the IMPF is 15.002 GHz with a relative standard deviation of 0.085%, and the fluctuation range of the center frequency is between 14.970 and 15.051 GHz. The maximum frequency drift is less than ± 0.051 GHz, yielding a 234-fold improvement in stability compared to the maximum frequency drift of up to 11.950 GHz when the MOMS-HC configuration is off. The average SF is 0.52 with a relative standard deviation of 0.667%, and the average bandwidth is 0.90 GHz with a relative standard deviation of 0.682%. During the 3-hour stability test, both the RR and roll-off rate remain above 29.89 dB and 16.57 dB/GHz, respectively. The detailed time-dependent variations of the SF, center frequency offset, bandwidth, and RR without and with the MOMS-HC configuration are shown in Figures 5d-g, respectively. The normal statistical distributions in the corresponding marginal plots indicate that the IMPF continuously stabilizes to the initial state. These results demonstrate that the MOMS-HC strategy effectively suppresses performance degradation caused by environmental disturbances, allowing the IMPF to maintain sustained stability and consistency over an extended period.

We also evaluate the self-recovery capability of the intelligently configured IMPF under extreme environmental disturbances. In this evaluation, we let the IMPF operate for 30 minutes with MOMS-HC on. The initial state is the same as that of the 3-hour stability test, operating with a center frequency of 15.010 GHz and a bandwidth of 0.90 GHz, as indicated by the sky-blue curve in Figure 5h. To simulate extreme disturbances, independent voltage disturbances are simultaneously applied to all the nine control ports at the beginning of the 4th BOC. These disturbances are step-type biases with random polarity and random value, ranging from -200 to +200 mV. The red curve in Figure 5h illustrates the degraded frequency response of IMPF at the beginning of applying voltage disturbances to the control ports. We can observe that the center frequency of the IMPF instantaneously drifts toward higher frequencies by 1.637 GHz, and the RR decreases by 10.62 dB, from 30.27 dB to 19.65 dB. Concurrently, the SF drops to 0.14, and the passband is severely split. However, the MOMS-HC strategy enables the IMPF to recover to its initial operation state within 5 BOCs, as demonstrated in Figure 5h. During the subsequent 22-minute period, the IMPF maintains an optimal operation state with an RR exceeding 30.01 dB and a roll-off rate above 15.00 dB/GHz. Figure 5i indicates the waterfall diagram characterizing the stable performance of the IMPF from the 9th to the 30th minute. Meanwhile, Figures 5j-m display the temporal variations of SF, frequency offset, bandwidth, and RR over the 30-minute evaluation period, respectively.

The reconfigurability and stability of IMPFs directly determine their flexibility and reliability in practical RF communication tasks. Meanwhile, other key performance indicators of IMPFs, including RR, in-band ripple, and SF, remain crucial. Benefiting from the hardware-software synergy between the CROW device and the MOMS-HC configuration strategy, our proposed IMPF has achieved exceptional performance synergy.

### 2.5 IMPF-based channel equalization

In high-speed and large-capacity communication systems, signals using advanced modulation formats are prone to distortion induced by electro-optic and photoelectric conversion, electrical amplifiers, and fiber dispersion. Among these challenges, frequency-dependent fading stands as the primary impairment mechanism. This is primarily caused by two factors: the chromatic dispersion in optical fibers and the inherent frequency-dependent amplitude responses of the components (e.g., PM, PD, and EA) in system[50]. Traditionally, this issue can be solved by modifying the architecture of intensity modulation and direct detection systems, including single-sideband or vestigial-sideband modulation, intensity modulation with coherent detection, and Kramers-Kronig receiver[51,52]. However, these approaches inevitably increase system complexity and cost. Alternatively, advanced digital signal processing (DSP) techniques are employed to perform pre-equalization based on accurate channel estimation and channel feedback, but these techniques often suffer from high computational complexity and power consumption[53,54].

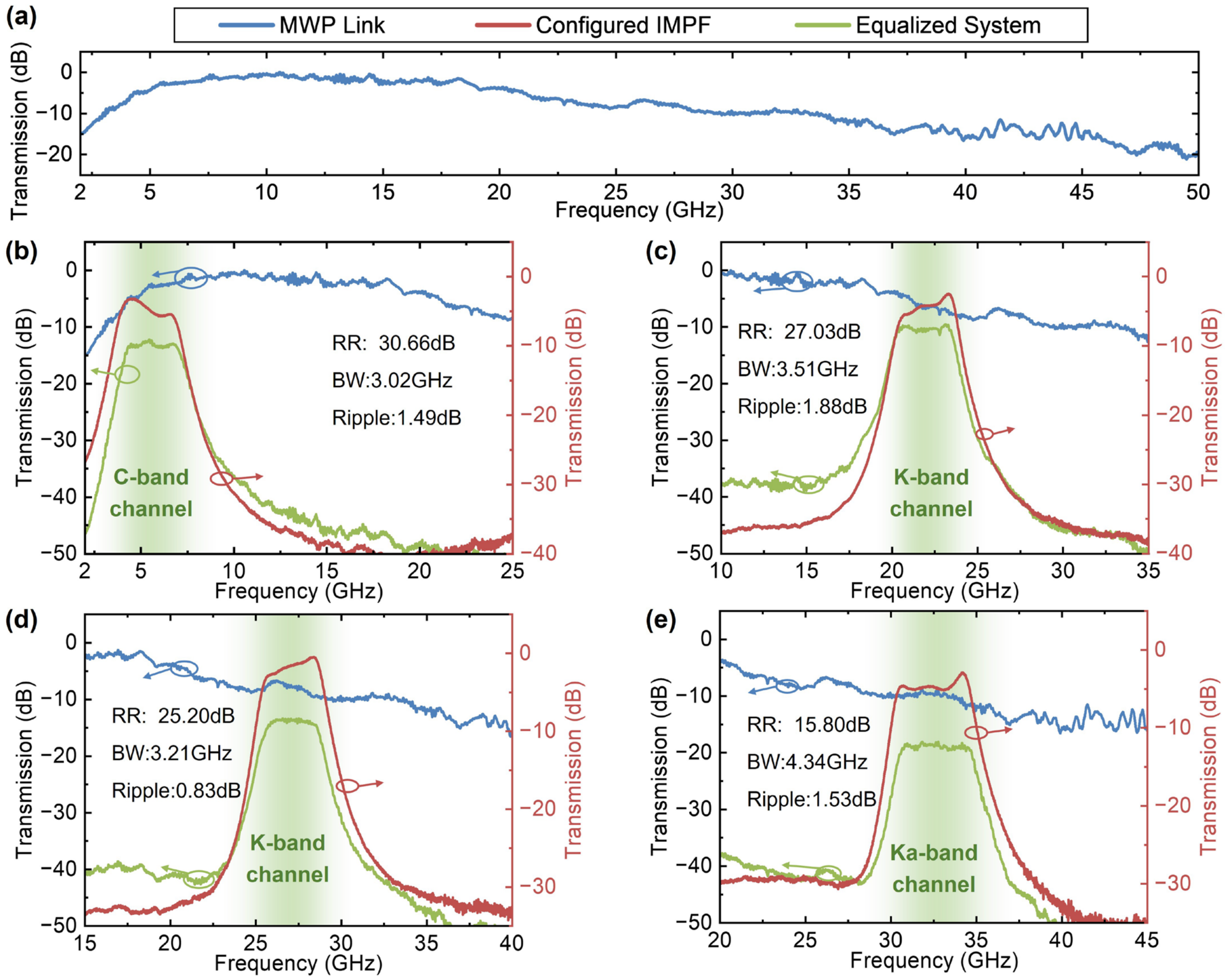

**Figure 6.** Channel equalization performance. (a) The amplitude frequency responses of a general microwave photonic system without the IMPF based on the 4th-order CROW. (b)-(e) Channel equalization results when the center frequency is (b) 5.5, (c) 22.0, (d) 27.0, and (e) 32.5 GHz, respectively. Green and red shaded areas denote the passband and stopband, respectively.

In this section, the intelligently configured IMPF, based on the MOMS-HC strategy, is employed to compensate for frequency-dependent fading and achieve channel equalization. This equalization capability is achieved by dynamically reshaping the passband response of the IMPF (see Supplementary Note II for more details). To demonstrate this capability, we directly utilize the microwave photonic system shown in Figure 3 to simulate the practical scenario with frequency-dependent fading. Figure 6a shows the inherent amplitude frequency response of the microwave photonic system without the IMPF based on the 4th-order CROW. We can observe significant amplitude ripples and degradation across the operational band from 2 to 50 GHz. These effects are attributed to the combined influence of the inherent responses of each device in the microwave photonic system. Specifically, within the 2–10 GHz range, the filtering edge of the OBPF adopted for SSB modulation induces higher loss for the lower frequency compared to the higher frequency. In the 10–50 GHz range, the combined effects of parasitic capacitance in the EA[55,56] and the bandwidth constraints of both the PD and PM lead to degradation in response flatness and insufficient responsivity at the high frequencies.

Table 2. Results of Channel Equalization

| Center (GHz) | BW[a)] (GHz) | Unequalized Ripple (dB) | Equalized Ripple (dB) | Ripple Reduction (dB) |
|---|---|---|---|---|
| 5.5 | 3.02 | 3.66 | 1.49 | 2.42 |
| 22.0 | 3.51 | 4.24 | 1.87 | 2.37 |
| 27.0 | 3.22 | 2.98 | 0.84 | 2.14 |
| 32.5 | 4.34 | 3.44 | 1.53 | 1.91 |

[a)] BW: bandwidth.

Therefore, we set the channel window centers at 5.5, 22.0, 27.0, and 32.5 GHz and set different target bandwidths for each to evaluate the dynamic equalization capability of the IMPF. The green and red curves in Figures 6b-e indicate the amplitude frequency responses of the configured IMPF and the equalized system, respectively. We can see that multiple equalized microwave photonic systems with flat passbands at various center frequencies are established. These responses demonstrate that, owing to the MOMS-HC strategy, the IMPF passband is adjusted to match the inherent response of the system, and its transition edges are also adjusted to ensure excellent symmetry in the equalized system response. Table 2 shows channel equalization results with different center frequencies and bandwidths. We can see that when the center frequency of the channel window is respective 5.5, 22.0, 27.0, and 32.5 GHz, the corresponding in-band ripple before equalization is 3.66, 4.24, 2.98, and 3.44 dB, respectively. After applying the IMPF-based equalization to the microwave photonic system, the corresponding in-band ripple is reduced to 1.49, 1.87, 0.84, and 1.53 dB, respectively. Therefore, these experimental results demonstrate the effectiveness of IMPF-based equalization. The overall RF gains of the IMPF system are -12.23 dB, -9.50 dB, -13.38 dB, and -18.10 dB, corresponding to different channel windows, as shown by the peak values of the green curves in Figures 6b-e, respectively. The key factors influencing RF gain include laser power, the half-wave voltage of the PM, the responsivity of the PD, the gains of the EDFA and EA, and the insertion losses of both the fabricated device and the OBPF used for SSB modulation.

Additionally, we can observe from Figures 6b-e that the response of the equalized system has different excess ILs at different channel windows. This is because the PM used in the experiment is polarization-dependent, and SOP variations during the configuration process can result in the system's gain degradation. To solve this problem, an electrically controlled PC can be used and co-configured by the MOMS-HC strategy[57].

## 3. Discussion

Based on our proposed MOMS-HC strategy, we have successfully implemented an intelligently configured IMPF with continuous tuning of the center frequency from 2 to 48 GHz and a wide bandwidth reconfiguration range from 0.66 to 4.15 GHz, featuring self-stabilization and self-recovery capabilities. The MOMS-HC strategy also significantly enhances the performance consistency of IMPF during the frequency tuning. In contrast to conventional equalization techniques, the MOMS-HC strategy enables the IMPF to equalize intra-channel fading through direct passband reshaping, eliminating the need for extra costly devices or digital equalization algorithms. This hardware-instead-of-software approach promises to enable end-to-end integrated microwave photonic system architectures[7,42].

Meanwhile, the intelligent configuration process of the IMPF is fundamentally an in-situ training process. The "train-as-deploy" paradigm ensures that the IMPF can be directly used for specific wireless communication tasks once the training is complete. Owing to this paradigm, the IMPF also circumvents the reliance on electromagnetic simulations and the challenges of physical alignment faced by traditional methods. In addition to IMPFs, the MOMS-HC strategy is also universally applicable to IOPUs of arbitrary architectures and can be extended to any scenario where a FOM can be defined to evaluate the system's state.

Notably, bandwidth reconfiguration of IMPFs over a broader range remains challenging, particularly for achieving bandwidths as narrow as the MHz level. Two main factors limit the reconfigurable bandwidth range of the CROW-based IMPFs. Firstly, high waveguide loss results in an inadequate microring Q-factor, which constrains the lower limit of the reconfigurable bandwidth. The utilization of optimized multimode waveguides with Euler bends[18,37] or free-form curved[58] microrings can overcome this limitation while improving the narrowband filtering performance of the IMPF. Secondly, the combined effects of the non-ideal DC self-coupling coefficient and thermal crosstalk limit the adjustable range of the equivalent coupling coefficient. This limitation consequently reduces the maximum reconfigurable bandwidth and hinders arbitrary passband shaping. Increasing the interval between the two arms of the MZI coupler and the length of the two arms[23,26], as well as carving grooves between microheaters[15], can enlarge the adjustable range of the MZI coupling coefficient. Additionally, employing higher-order CROWs can also further expand the reconfigurable bandwidth. Meanwhile, higher-order CROWs feature more flexible and adaptable passband response characteristics, enabling superior compensation for complex intra-channel fading.

Furthermore, the packaged chip in this work lacks TEC integration, leading to potential overvoltage and excessive power consumption. Future integration of TEC modules can address these limitations through active temperature stabilization and thermal management[12,26].

## 4. Summary

In summary, we have demonstrated an intelligently configured bandpass IMPF by employing the MOMS-HC strategy. The high-performance IMPF features four key capabilities: wideband center frequency tunability, flexible bandwidth reconfigurability, self-stabilization, and excellent channel equalization. The roll-off rate, RR, and SF of the IMPF reach up to 17.50 dB/GHz, 37.67 dB, and 0.78, respectively. Notably, with the MOMS-HC strategy, the IMPF can operate stably for 3 hours, and its frequency stability is improved by 234 times. Furthermore, the MOMS-HC strategy also enables the IMPF to recover to its target operating state quickly when all control ports experience significant disturbances simultaneously, which is rather challenging for manual manipulation. The MOMS-HC strategy also enables dynamic compensation of intra-channel fading through reshaping the passband response of the IMPF, thus achieving channel equalization. This capability positions the intelligently configured IMPF as a viable alternative to conventional pre-equalization achieved by DSP. Overall, this work makes a solid step toward applications of the IMPF in practical microwave systems.

## 5. Methods

### 5.1 Design of the MOMS-HC strategy

The MOMS-HC strategy is a task-driven intelligent configuration framework with four stages. Its detailed procedure is as follows. The execution flowchart of the MOMS-HC strategy, which consists of four stages, is shown in Figure M1. In the "Preprocessing Stage" (Stage I), the MOMS-HC strategy first calculates and defines various properties of the IMPF programmatically. The FOM is then predefined for the subsequent configuration. Considering the global search capability and fast convergence of the PSO algorithm[46], the MOMS-HC strategy selects PSO for rapid global optimization in the "Global Search Stage" (Stage II). The PSO algorithm randomly generates an initial state, where each particle is represented as a multi-dimensional vector corresponding to the multiple control voltages one-to-one. The fitness of each particle (i.e., the FOM) is defined as a function of the most important property of the IMPF to achieve rapid global optimization. In the absence of a packaged thermo-electric cooler (TEC), there is a strong correlation between the chip temperature and the magnitude of initial control voltages, which makes the configuration of the IMPF sensitive to the initial point. Therefore, the MOMS-HC strategy iterates Stage II over multiple random initial points until the convergence criteria are met, and the optimal state is selected at the maximum FOM.

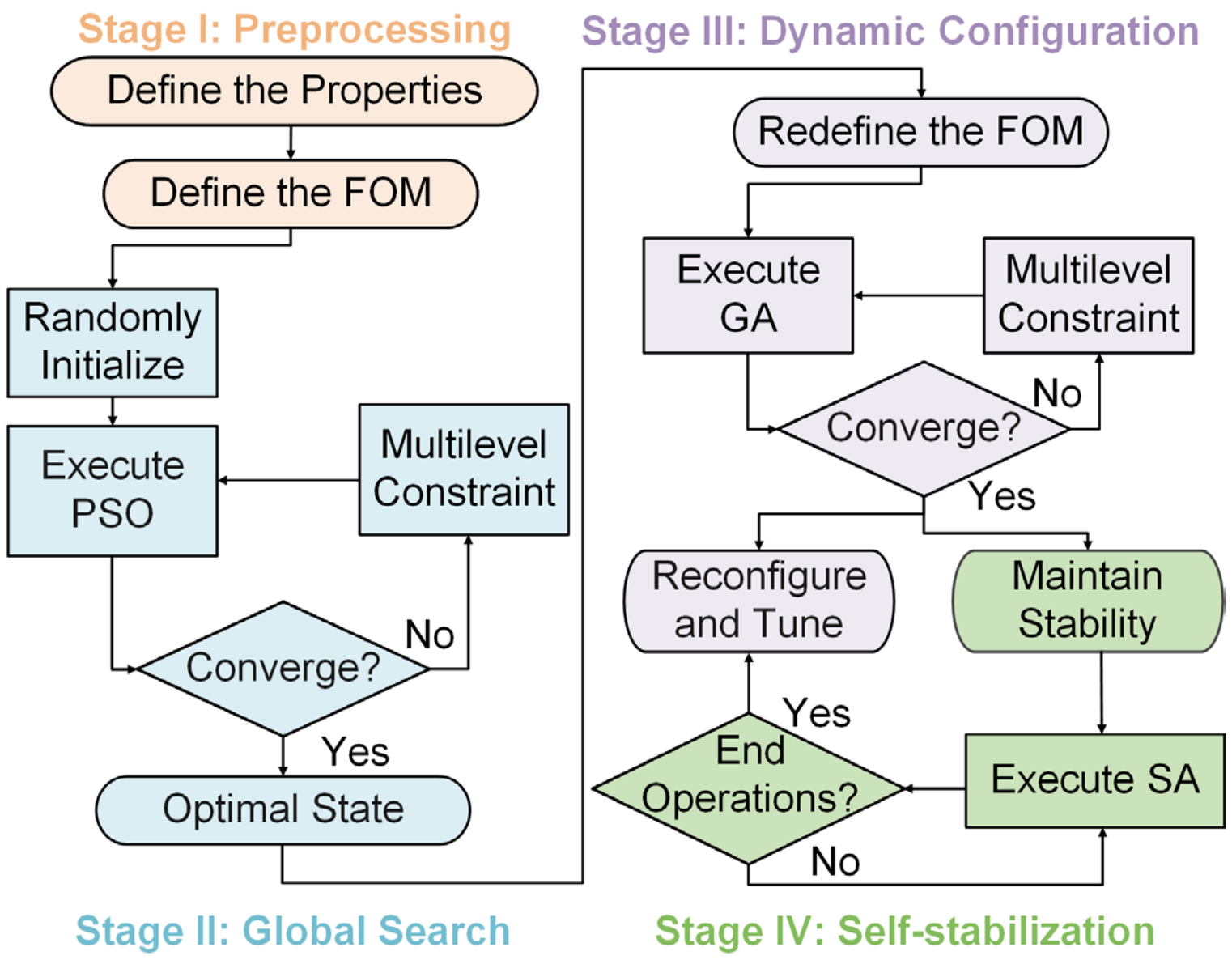

**Figure M1.** The flowchart of the proposed MOMS-HC strategy.

The core of the "Dynamic Configuration Stage" (Stage III) is a swarm intelligence optimizer based on the standard GA[47], which specializes in precise optimization under multi-objective constraints. The GA performs collaborative multi-objective optimization with a redefinable FOM. Each individual consists of multiple genes, and the initial genes are obtained from Stage II. The population size should be set to a relatively large value, typically more than 20 times the number of parameters, to balance the global search capability and rapid convergence speed. Based on standard genetic operations, such as roulette-wheel selection, gene crossover, mutation, and environmental selection, the MOMS-HC strategy can balance the weights of different objectives to avoid local optima in multi-objective optimization problems and maintain excellent search capability in uncertain environments, thus enabling IMPF to achieve synergy of multiple properties. Notably, the global search of control voltage significantly exacerbates the overall temperature fluctuation of the chip due to electric power variations and crosstalk, leading to serious degradation of IMPF performance. Therefore, in Stage II and Stage III, the MOMS-HC strategy incorporates a multi-level constraint mechanism on the voltage search space, effectively suppressing the overall temperature fluctuations.

Following the completion of Stage III, the "Self-stabilization Stage" (Stage IV) will be initiated to ensure sustained and stable operation of IMPF. Alternatively, iterative re-execution of Stage III enables dynamically adjusting the operational properties of IMPF, including bandwidth reconfiguration and center frequency tuning. Considering the environmental disturbances in practical operating environments, such as temperature fluctuations[43], mechanical vibrations, and power supply noise[44,45], the control voltage must be dynamically adjusted to compensate for the performance degradation of the IMPF. Therefore, in Stage IV, the MOMS-HC strategy performs a local fine-tuning search using the SA algorithm[48]. Through a customized temperature decay mechanism, SA enables rapid local optimization while progressively constraining the amplitude of voltage adjustments. Initially, the high annealing temperature permitted a relatively large voltage search range. Later, the annealing temperature rapidly reduces to restrict the voltage search range, thereby suppressing secondary performance fluctuations induced by voltage search and ensuring the performance consistency and self-stabilization of the IMPF under complex operating conditions (see Supplementary Note I for more details of the sub-algorithm).

**5.2 Precision and time cost of the configuration**

In the 3-hour stability test, when the MOMS-HC strategy is applied to the IMPF, its center frequency drifts between +0.051 GHz and -0.030 GHz relative to the target value of 15.000 GHz, and its bandwidth varies between +0.02 GHz and -0.02 GHz relative to the target value of 0.90 GHz. These deviations are primarily attributed to the combined effects of optimization constraints imposed by the BOC setting and measurement errors induced by experimental setup limitations, such as the 0.010 GHz resolution of the VNA.

In our experiment, the sensors (OSA and VNA) and actuator (VSA) are connected to the controller (computer) via Ethernet ports utilizing the TCP/IP protocol and serial ports, resulting in a total communication speed of approximately 2 Hz. Without relying on any prior experience, the first configuration of IMPF can be finished within 2000 s (less than 40 iterations, with 100 calculations per iteration), while the response time of the TiN heater is approximately 10 μs[59]. For center frequency tuning with the same bandwidth or bandwidth reconfiguration at the same center frequency, the MOMS-HC strategy requires only a single execution of Stage II. For example, after configuring the IMPF to a specific response (e.g., the center frequency is 25 GHz and the bandwidth is 0.66 GHz), the strategy can directly use the current control voltages as the starting point and repeat Stage III to reconfigure the IMPF. This mechanism significantly reduces the time cost. Therefore, the single reconfiguration of the IMPF, including bandwidth reconfiguration and center frequency tuning, can be completed within 400 s (less than 8 iterations). If the response speed of the sensors and actuators is increased to make the component communication time cost negligible, the first configuration and the single reconfiguration of IMPF can be completed in 40 ms and 8 ms, respectively. Therefore, to further reduce the time cost, techniques such as optoelectronic co-integration[60] and frequency-to-time mapping[61] can be used in the future.

**Conflict of interest**

The authors declare no competing financial interest.

**Data Availability Statement**

The data that support the findings of this study are available from the corresponding author upon request.

**Acknowledgements**

This project was supported by the National Key Research and Development Program of China (2024YFF0728000), the Program for HUST Academic Frontier Youth Team (2018QYTD08), and the Project of Key Laboratory of Radar Imaging and Microwave Photonics (Nanjing University of Aeronautics and Astronautics), Ministry of Education (NJ20230001).

**Supporting Information**

Supporting Information is available from the Wiley Online Library.

Supplementary Information for

**Intelligent Configuration of Integrated Microwave Photonic Filter Featuring Self-Stabilization and Programmable Response**

*Yutong Shi, Yuan Yu**, *Yifan Liu, Kaixiang Cao, Mengmeng Deng, Fangzheng Zhang, Hailong Zhou, and Xinliang Zhang**

**Correspondence E-mail:* yuan_yu@hust.edu.cn; xlzhang@mail.hust.edu.cn

# Contents

**Supplementary Note I: Details of the optimization sub-algorithm**

*Particle swarm optimization algorithm* is a population-based algorithm. Each individual (particle) of the PSO algorithm has position and velocity attributes within the search space. Its movement is guided jointly by the personal historical best solution ($pbest_i$) and the population global best solution ($gbest$). Particles perform searches via velocity-position updating mechanisms:

$$v_i^{k+1} = \omega v_i^k + c_1 r_1 (pbest_i - x_i^k) + c_2 r_2 (gbest - x_i^k), \quad \text{(S-1)}$$

and

$$x_i^{k+1} = x_i^k + v_i^{k+1}, \quad \text{(S-2)}$$

where the adaptive linear inertia range ($\omega$ varies from 0.1 to 1.1) balances the global and local search capabilities, and self-adjusting weight ($c_1$=1.5) and social-adjusting weight ($c_2$=2.0) control the individual experience and social information weights. The fitness function directly evaluates the value of the FOM corresponding to the particle position. PSO balances exploration and development through the group information-sharing mechanism and is classified as a global search algorithm.

*Genetic Algorithm*, also known as the Evolutionary Algorithm, simulates the mechanisms of natural selection and genetic variation. Individuals are represented through gene encoding, and their fitness is directly measured by the FOM. The population evolves iteratively through selection, crossover, and mutation operations. In this work, the selection operation retains individuals with high fitness, and the crossover operation recombines genes to explore new solutions (crossover rate=0.8). The uniform mutation operation introduces random perturbations to avoid premature convergence (mutation rate=0.01). The mutation intensity decays exponentially with generations to balance early global exploration and later local optimization.

*Simulated Annealing algorithm* is inspired by the annealing process in metal heat treatment and achieves optimization by simulating the physical mechanism of the slowly cooling high-temperature materials. In this algorithm, the individual (current solution) is associated with a fitness function (FOM) characterizing the energy of the system. The algorithm employs a temperature parameter to regulate the search process. During high-temperature phase, solutions with degraded objective values are accepted to enable global exploration. The subsequent cooling then shifts the search focus toward local refinement. New solutions are generated by random perturbations, and the decision to update the state is based on the Metropolis criterion. Overall, the SA algorithm simulates the two phases of the physical annealing process: high-temperature rapid search and low-temperature precise optimization. This mechanism aligns well with the ideal evolution process of thermo-optic tunable optical devices. In this work, we set the initial annealing temperature of SA to 110, enabling a broad search range. The exponential cooling function (ECF) is then applied to accelerate temperature reduction. We also set an exit mechanism for the algorithm. Once the IMPF achieves the target frequency response, the algorithm exits immediately. This not only saves computational overhead but also avoids unnecessary voltage searching that could cause temperature fluctuations.

**Supplementary Note II: FOM definitions of IMPF configuration**

During the configuration process based on the MOMS-HC algorithm, the FOM for each different algorithm stage is defined as follows.

*PSO*: In Stage II, the nine voltages are rapidly optimized to make the IMPF preliminarily approach the target response. We define $\mathrm{T(f)}$ as the transmittance at a specified frequency $\mathrm{f}$. To enhance the robustness of the global optimization in Stage II, we take a linear combination of four points in the 5 GHz range around the target center frequency as one item in the $\mathrm{FOM}_{PSO}$. The FOM is expressed by

$$\mathrm{FOM_{PSO}} = \left[T\left(\mathrm{center_{target}}\right) + \sum_{i=0}^{4} T\left(\mathrm{center}_i\right)\right] \cdot \frac{1}{5}, \tag{S-3}$$

*GA*: In Stage III, the configuration of the nine voltages is refined according to multi-objective optimization requirements. The definition of FOM comprehensively considers all properties, and is expressed by

$$\mathrm{FOM_{GA}} = \exp\left(-\lambda_1 \left|\frac{\Delta \mathrm{f_{center}}}{\mathrm{f_{target}}}\right|^2\right) \cdot \exp\left(-\lambda_2 \left|\frac{\Delta \mathrm{BW_{FWHM}}}{\mathrm{BW_{target_FWHM}}}\right|^2\right) \cdot \exp\left(-\lambda_3 \left|\frac{\Delta \mathrm{BW_{10dB}}}{\mathrm{BW_{target_10dB}}}\right|^2\right)$$
$$\cdot \exp\left(-\lambda_4 \cdot \mathrm{EF_{sym}} \cdot \mathrm{EF_{flat}}\right) \cdot \frac{\mathrm{RR}}{\mathrm{RR_{target}}} \cdot \delta\left(\mathrm{Num_{peaks}} = 1\right), \tag{S-4}$$

where $\Delta \mathrm{f_{center}}$ denotes the deviation between the current and target the center frequencies, $\Delta \mathrm{BW_{FWHM}}$ denotes the deviation between the current and target FWHM bandwidths, $\Delta \mathrm{BW_{10dB}}$ denotes the deviation between the current and target 10-dB bandwidths, $\lambda_\mathrm{i}$ $(\mathrm{i} = 1,2,3,4)$ denotes the i-th optimized intensity coefficient, and $\mathrm{EF_{symmetry}}$ and $\mathrm{EF_{flatness}}$ denote evaluation factors for the symmetry and flatness of the IMPF response, respectively. To ensure high-performance filtering in the IMPF, we impose a strict constraint via an indicator function $\delta\left(\mathrm{Num}_{peaks} = 1\right)$ to guarantee that the amplitude frequency response contains only one single peak. All parameters have been normalized to eliminate dimensional discrepancies.

*SA*: In the Stage IV, the nine voltages need to be fine-tuned dynamically to offset thermal drift and environmental disturbances. The FOM combines the degradation of multiple core properties, expressed by

$$\text{FOM}_{SA} = \frac{1}{1+\prod_{k=1}^{4}(1+\alpha_k|\Delta_k|)}\cdot\exp\left(-\beta(\text{Num}_{\text{peaks}}\text{-}1)^2\right), \tag{S-5}$$

where $\Delta_1$, $\Delta_2$, $\Delta_3$, and $\Delta_4$ denote the deviations of center frequency, FWHM bandwidth, shape factor (SF, defined as the ratio of 3-dB to 10-dB bandwidth), and rejection ratio (RR), respectively. Meanwhile, $\beta$ denotes penalty intensity for peaks count ($\beta$>10), and $\alpha_k$ denotes a normalization factor, respectively. The multiplicative form of the FOM ensures that the degradation in any one property will significantly reduce the overall FOM. This mathematical interdependence requires real-time synchronized optimization of all four key properties to ensure stable and sustained operation of the IMPF.

Based on the above FOM definition framework, configuring the IMPF only needs to determine the target and current values of key properties according to specific requirements. Notably, the frequency response reshaping of the IMPF primarily relies on two methods to define these values. The first method extracts the responses of IMPF in multiple specific frequency bands based on the targets of response reshaping, thereby determining the target and current values of key properties from the sliced response. The second method employs the frequency response of the microwave photonic (MWP) link with the 4th-order CROW as the configuration object and directly determines these values according to equalization targets.

**Supplementary Note III: Procedure and challenges of manual manipulation**

The operator can also manually manipulate the control voltages to achieve the desired response. Based on our experimental experience, the specific procedure for manipulating the 4th-order CROW-based IMPF is as follows. Firstly, the voltages applied to $H_1$-$H_4$ are adjusted with large steps to roughly align the resonant wavelengths of the four microrings. Subsequently, the voltages applied to $H_5$-$H_9$ are adjusted to optimize the filter shape by modifying coupling coefficients. Experiments indicate that $H_7$ exhibits the most significant influence on device response, owing to thermal crosstalk and coupling dependence, followed by $H_6$ and $H_8$. Notably, adjusting either microheater of MZI-based coupler not only modifies the coupling coefficient but also alters resonant wavelengths of MRRs due to simultaneous optical phase shift. Moreover, the thermal crosstalk can further exacerbate this inter-coupling between coupling coefficient and resonant wavelengths. Therefore, the voltages applied to $H_1$-$H_4$ need to be readjusted for compensation. Meanwhile, resonant wavelength adjustments also influence the coupling coefficients of MRRs due to optical phase shift and thermal crosstalk. This bidirectional coupling relationship forces the operator to repeatedly adjust the resonant wavelength and the coupling coefficient alternately to find the dynamic balance, which is very challenging for manual manipulation.

**Supplementary Note IV: Measured optical spectra of modulated signals**

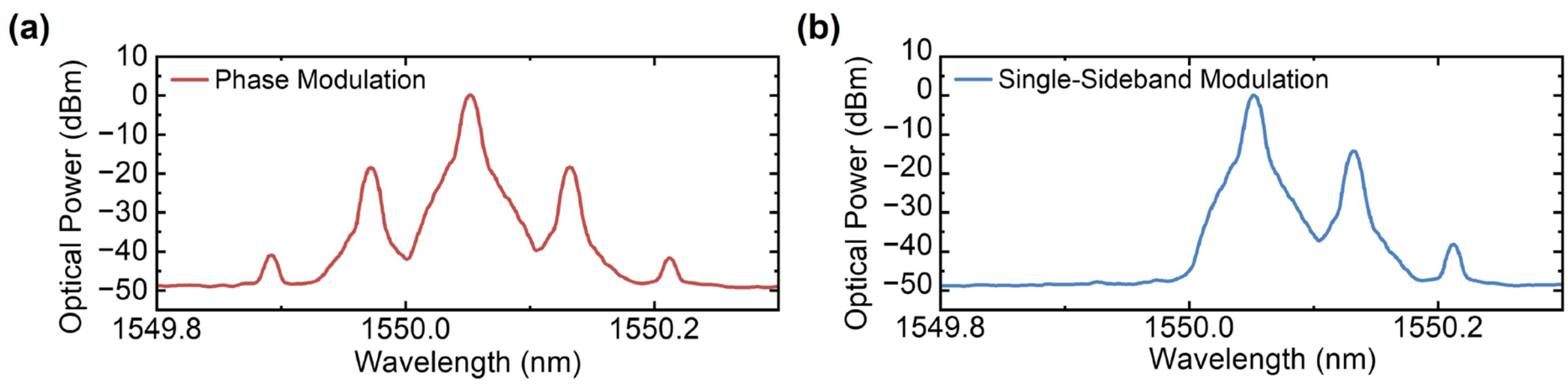

**Figure S1.** Optical spectra of (a) the phase-modulated signal, and (b) the SSB signal at a modulation frequency of 10.0 GHz.

Figure S1a and S1b show the measured optical spectra of the phase-modulated signal and the SSB signal after OBPF when the modulation frequency is 10 GHz, respectively. The wavelength of the optical carrier is 1550.05 nm. The SSB modulation is used to maximize the tuning range of the microwave photonic filter.

**Supplementary Note V: Intelligent configuration of DUT with MOMS-HO algorithm**

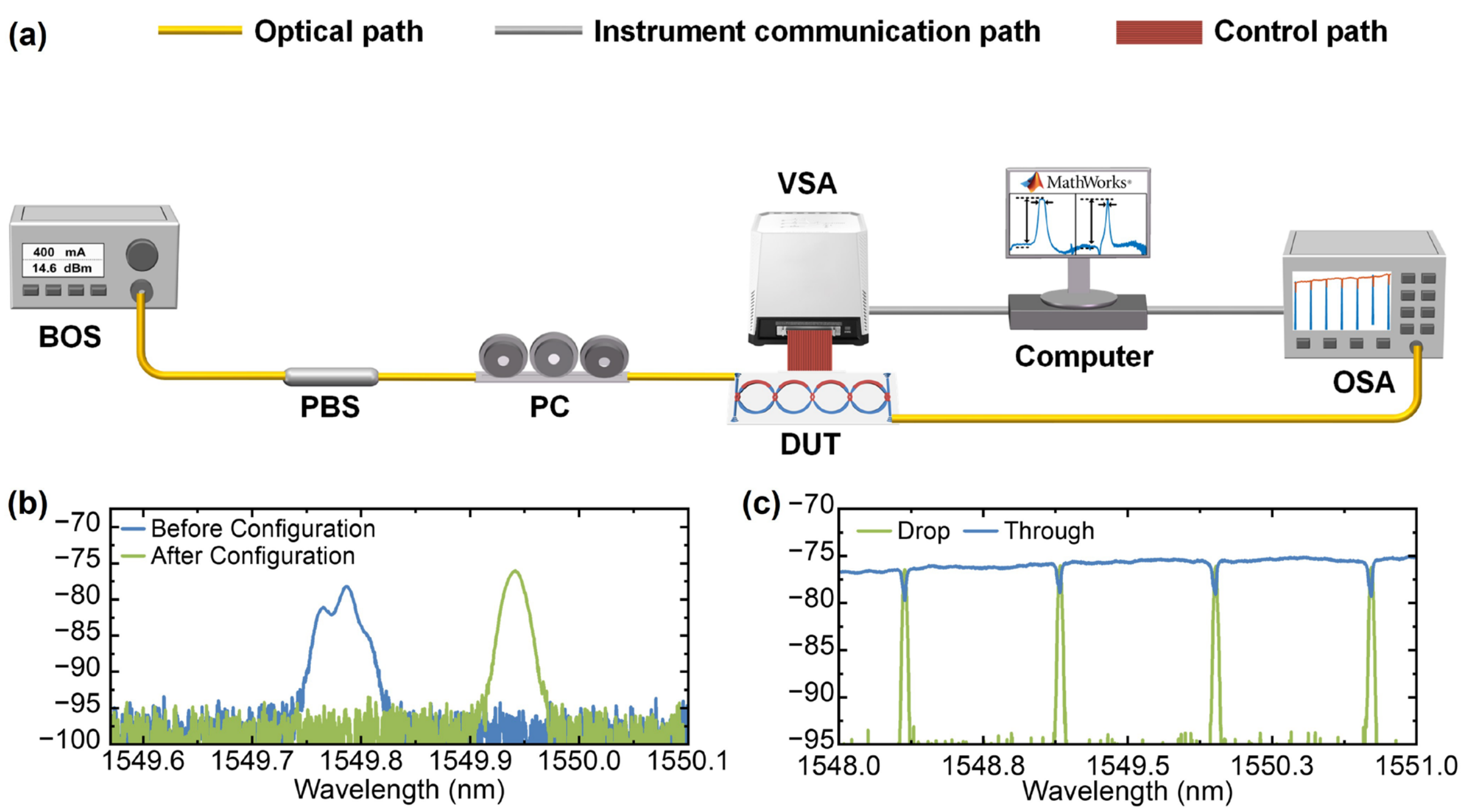

**Figure S2.** (a) Experimental setup for measuring the optical transmission spectrum of DUT. (b) Optical transmission spectra of the 4th-order CROW at the output port before configuration (blue curve) and after configuration (green curve). (c) Measured optical transmission spectra of the 4th-order CROW from GC1 to GC4 (through port, blue curve) and GC3 (drop port, green curve). BOS broadband optical source, PBS polarization beam splitter, PC polarization controller, DUT, device under test, VSA, voltage source array, OSA optical

spectrum analyzer.

To measure the optical spectrum across a broader wavelength range and extract the free spectral range (FSR) of the device under test (DUT), we first characterize the optical transmission spectrum of the DUT using the experimental setup shown in Figure S2a. The broadband light emitted from a broadband optical source (BOS) is first injected into a polarization beam splitter (PBS) to obtain linearly polarized light. Then, the state of polarization (SOP) is adjusted to be aligned with that of the grating couplers ($GC_1$) of the DUT using a polarization controller (PC). Finally, the processed optical signal is coupled out from $GC_3$ and received by an optical spectrum analyzer (OSA, YOKOGAWA AQ6319).

When all the voltages applied to the microheaters by the voltage source array (VSA) are 0 V, the measured optical transmission spectrum of DUT is shown as the blue curve in Figure S2b. We can see that the passband of DUT is not flat and even splits into two peaks. This is caused by misalignment of the MRR resonant wavelengths and coupling coefficients, both resulting from fabrication errors.

Considering that manual manipulation cannot rapidly and efficiently configure these multiple parameters to attain optimal solutions (see Supplementary Note III), we directly demonstrate the intelligent configuration of DUT based on the MOMS-HC algorithm. The target response of DUT is set to be an optical bandpass filter with a bandwidth of 0.016 nm (2 GHz) and an RR higher than 15 dB. The measured optical transmission spectrum of DUT is shown as the green curve in Figure S2b, indicating that an optical filter with an RR of 18.93 dB and an FWHM bandwidth of 0.018 nm (2.25 GHz) is realized. We can observe that the configured FWHM bandwidth deviates from the target by as much as 0.25 GHz. This is because the resolution of OSA is only 0.02 nm, which cannot precisely characterize the FWHM bandwidth of DUT. By contrast, the vector network analyzer (VNA) can achieve a much higher resolution than the OSA. Therefore, accurate configuration, measurement, and characterization of the IMPF will be performed using the experimental setup shown in Figure 3.

Figure S2c illustrates the optical transmission spectra from $GC_1$ to $GC_4$ (through port, blue curve) and from $GC_1$ to $GC_3$ (drop port, green curve) over multiple FSRs. We can see that the measured FSR of DUT is 0.8 nm (100 GHz).

**Supplementary Note VI: Intelligent configuration of the IMPF**

We also demonstrate an intelligent configuration of the IMPF based on the 4th-order CROW, aiming for a center frequency of 30 GHz, an FWHM bandwidth of 3 GHz, an RR higher than 30 dB, an SF higher than 0.7, and an in-band ripple smaller than 1.0 dB. Figure S3a shows the normalized FOM as a function of the number of iterations during the configuration process, where the blue and green curves correspond to Stage II and III, respectively (see Supplementary Note II for FOM definitions). The preprocessing and predefined steps in Stage I are omitted here due to their negligible time cost. Figure S3b-e and S3f-i show the voltage distribution applied to all microheaters from $H_1$ to $H_9$ and the amplitude frequency response of the IMPF after different iterations, corresponding to nodes "I", "II", "IV", and "V", respectively. Specifically, Figure S3b and S3f present the results after 8 iterations of the PSO algorithm in Stage II, corresponding to node "I" in Figure S3a. At this node, the IMPF demonstrates a basic filtering function, but the passband exhibits splitting with poor edge symmetry, and the center frequency is approximately 40 GHz, as shown in Figure S3f. This degradation is due to the significant temperature fluctuations of the chip caused by the extensive voltage search in this stage. However, the degradation does not affect the subsequent configuration.

In Stage III, the normalized FOM increases as the number of iterations increases. However, it does not increase monotonically and exhibits significant degradation and oscillation in early iterations, as shown by the FOM variation between node "I" and node "III" in Figure S3a. There are four main reasons for this phenomenon. Firstly, the global search for optimal control voltage at the beginning of each iteration leads to fluctuations in chip temperature. This is because abrupt voltage jumps (e.g., sudden high-to-low transitions) induce drastic heating power variations, which consequently lead to chip temperature fluctuations. Secondly, the adjustable parameters of the CROW-based IMPF interact with each other due to the complexity of its inherent structure. For example, adjusting either microheater of the MZI-based coupler not only modifies the coupling coefficient between the two adjacent MRRs but also alters their resonant wavelengths due to simultaneous optical phase shift. Moreover, the thermal crosstalk can further exacerbate the inter-coupling between these thermo-optic adjustable parameters. Thirdly, GA needs to simultaneously handle multiple optimization objectives, including FWHM bandwidth, center frequency, RR, SF, roll-off rate, insertion loss (IL), in-band ripple, response symmetry, and response smoothness associated with the absence of passband splitting, which results in a high degree of complexity in the evolution process. As a result, a certain amount of time is required for adaptive selection to balance the weights of different objectives. Fourthly, environmental temperature fluctuations and mechanical vibrations, as well as power supply noise, also lead to degradation of the IMPF performance. These factors are also the main challenges in achieving high-performance IMPF.

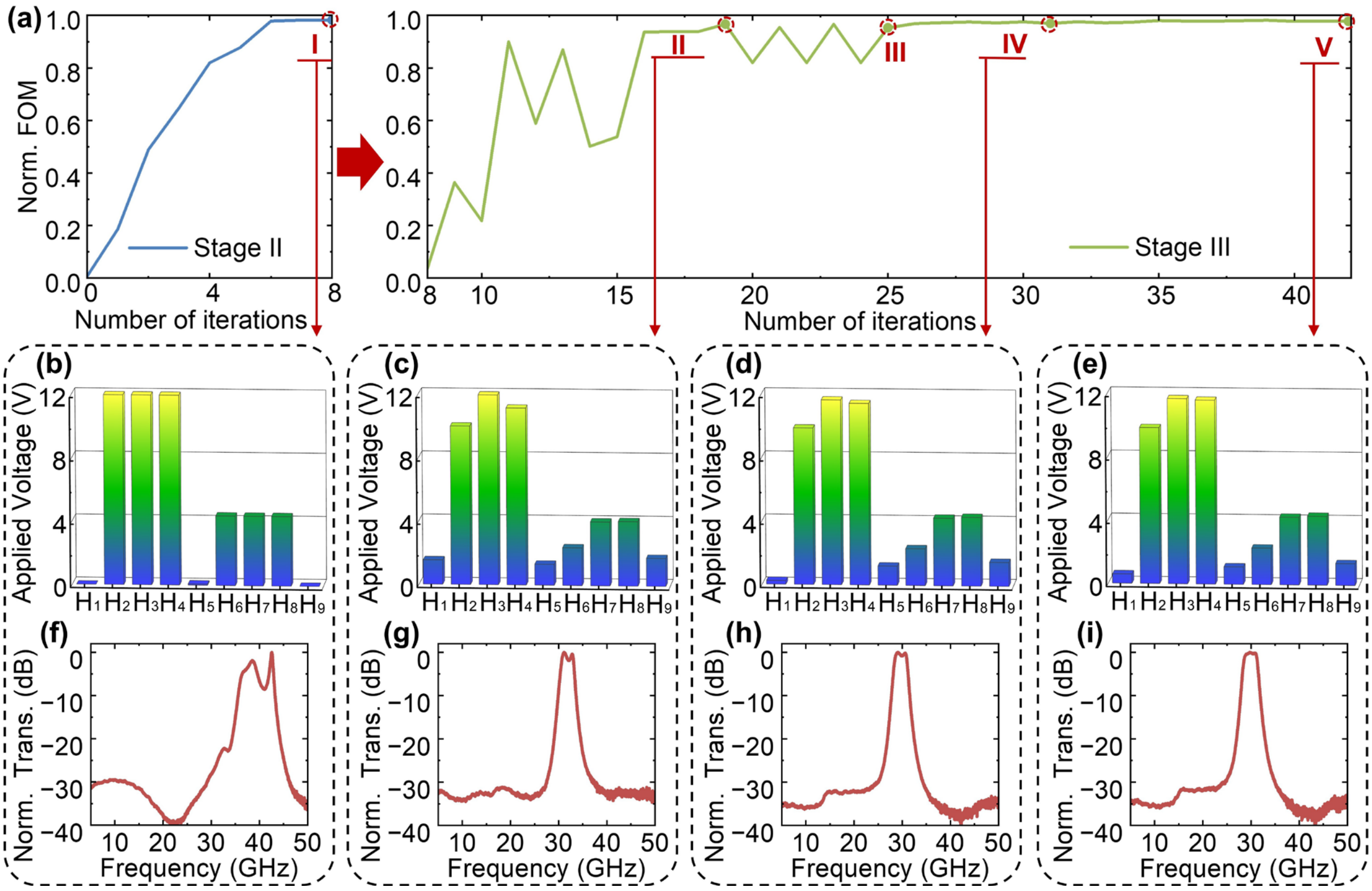

**Figure S3.** Intelligent configuration of the IMPF. (a) The variation of the normalized FOM as the number of iterations increases during the intelligent configuration process. (b)-(e) Voltage distribution applied to all the microheaters in the DUT at the (b) 8th, (c) 19th, (d) 31st, and (e) 42nd iterations, respectively. (f)-(i) Amplitude frequency response of the IMPF during the intelligent configuration process at the (f) 8th, (g) 19th, (h) 31st, and (i) 42nd iterations, respectively.

After 25 iterations, the FOM reaches 0.953, as shown by node "III" in Figure S3a. We can see that the frequency response of the IMPF gradually converges, indicating that the overall temperature of the chip is stabilizing. Figure S3h shows the frequency response of the IMPF after 31 iterations, which corresponds to node "IV" in Figure S3a. The normalized FOM reaches 0.969. The overall performance is satisfactory except for the in-band ripple, which is 0.85 dB. Subsequent iterations focus on further flattening the passband to approach the target response. Finally, after 42 iterations, the FOM reaches 0.982, corresponding to node "V" in Figure S3a. The corresponding IMPF achieves an FWHM bandwidth of 2.97 GHz, a center frequency of 29.899 GHz, an RR of 31.21 dB, an in-band ripple of 0.52 dB, an SF of 0.74, and a roll-off rate of 13.08 dB/GHz simultaneously, as shown in Figure S3i. The results show that all properties of the IMPF perfectly meet the expected targets.

**Supplementary Note VII: The rationale of the BOC setting**

In the self-stabilization test, the 60-second basic operating cycle (BOC) is set primarily based on two core factors: the response speed of sensors and actuators (e.g., VNA, VSA, and TiN microheaters) and the balance between operational stability, performance monitoring accuracy, and dynamic adjustment capability.

First, the BOC accounts for the measurement and control latency of the experimental system. As stated in the manuscript, the VNA acts as a sensor to monitor the IMPF's frequency response, while the VSA serves as an actuator to apply control signals. In our experiment, the total communication speed between the controller (computer) and these instruments is approximately 2 Hz. This means both data acquisition by the VNA and control signal transmission to the VSA require a certain amount of time. A 60-second BOC provides sufficient time for the accurate measurement and dynamic configuration of the IMPF's frequency response.

Second, during each BOC, SA dynamically adjusts the voltage of the control port and calculates the voltage compensation in real time to suppress environmental disturbances. When facing significant environmental disturbances, SA requires a certain amount of time for optimization iterations to ensure the operational stability of the IMPF. Additionally, the IMPF's frequency response needs to be captured at an appropriate time interval during the self-stabilization test. This enables systematic tracking and monitoring of the long-term evolution of IMPF's performance metrics, including center frequency drift, bandwidth, RR, and SF. Therefore, the BOC is set to 60 seconds to strike a balance between operational stability, performance monitoring accuracy, and dynamic adjustment capability.